\documentclass[10pt, twoside]{article}
\usepackage[utf8]{inputenc}
\usepackage[english]{babel}

\usepackage{hyperref}
\usepackage{graphicx}
\usepackage{wrapfig}
\usepackage{amsmath}
\DeclareMathOperator\arctanh{arctanh}

\usepackage{lipsum}
\usepackage{framed}
\usepackage{layout}
\usepackage{amsmath,amssymb}
\usepackage{bm}
\usepackage[sc]{mathpazo}
\usepackage[T1]{fontenc}
\usepackage{fancyhdr}
\usepackage{microtype}
\usepackage{multicol}
\usepackage{supertabular}
\usepackage{alltt}
\usepackage{siunitx}
\usepackage{float}
\usepackage{newfloat}
\DeclareFloatingEnvironment[name={Gráfico}]{graph}
\usepackage[section]{placeins}
\usepackage[usenames,dvipsnames]{color}
\usepackage{lettrine}
\usepackage{paralist}
\usepackage{multirow} 
\usepackage{booktabs} 
\usepackage{xfrac}
\usepackage{booktabs}
\usepackage{chngcntr} 
\usepackage{gensymb}
\usepackage{pdfpages}

\usepackage{mathtools}
\DeclarePairedDelimiter\abs{\lvert}{\rvert}
\DeclarePairedDelimiter\norm{\lVert}{\rVert}
\makeatletter
\let\oldabs\abs
\def\abs{\@ifstar{\oldabs}{\oldabs*}}
\makeatother
\let\oldnorm\norm
\def\norm{\@ifstar{\oldnorm}{\oldnorm*}}
\makeatother

\usepackage{subcaption}

\usepackage{titlesec}
\renewcommand\thesection{\Roman{section} \centering}
\renewcommand\thesubsection{\arabic{subsection}}

\titleformat{\section}[block]{\normalsize\scshape\filcenter}{\thesection.}{1em}{}
\titleformat{\subsection}[block]{\normalsize\scshape\filcenter}{\thesubsection.}{1em}{}

\usepackage[left=3.5cm, right=3.5cm, top=2.5cm,bottom=3cm]{geometry}

\usepackage{lmodern}

\title{\large \textbf{RIEMANNIAN MANIFOLDS DUAL TO STATIC SPACETIMES}}

\author{
Carolina Figueiredo and Jos\'{e} Nat\'{a}rio \\
{\small CAMGSD, Departamento de Matem\'{a}tica, Instituto Superior T\'{e}cnico,}\\
{\small Universidade de Lisboa, Portugal}
\vspace{1mm}
}

\date{}

\begin{document}

\maketitle

\textsc{Abstract.} We establish a one-to-one correspondence between static spacetimes and Riemannian manifolds that maps causal geodesics to geodesics, as suggested by L.~C.~Epstein. We explore constant curvature spacetimes -- such as the de Sitter and the anti-de Sitter spacetimes -- and find that they map to constant curvature Riemannian manifolds, namely the Euclidean space, the sphere and the hyperbolic space. By imposing the conditions required to map to the sphere, we obtain the metrics for which there is radial oscillatory motion with a period independent of the amplitude. We then consider the case of a perfect fluid and an Einstein cluster and determine the conditions required to find this type of motion. Finally, we give examples of surfaces corresponding to certain types of motion for metrics that do not exhibit constant curvature, such as the Schwarzschild, Schwarzschild de Sitter and Schwarzschild anti-de Sitter solutions, and even for a simplified model of a wormhole.

\section*{Introduction}

The notion of curvature of Riemannian manifolds is intuitive and can be easily understood with elementary mathematical concepts. However, when it comes to curved spacetimes, this notion becomes considerably less clear. Notwithstanding, the motion of test bodies under the action of a gravitational field, given by the geodesics of the corresponding spacetime, can only be properly explained by the warping of time resulting from the spacetime curvature. As for the concept of geodesic, a curve of extremal length between two points of space, it also becomes less intuitive when working with spacetimes. A Riemannian manifold, with its definite positive metric, allows us to measure distances, and thus lengths of curves, as we do naturally with a ruler. Under these circumstances, the concept of geodesic can be easily understood. By opposition, the Lorentzian signature of spacetimes leads to curves with zero length, and thus distances no longer match our intuitive notion. This makes it difficult to visualize spacetime geodesics, and consequently to decode the possible types of motion.

There have been many approaches to overcome these difficulties. In 1981, diSessa \cite{DiSessa} proposed a \textit{``map-making/wedgie calculus''} approach to track the geodesics of spherically symmetric spacetimes. Marolf \cite{Marolf} used an embedding into a $(2+1)$-dimensional Minkowski spacetime, called an \textit{``embedding diagram''}, to decode the features of the radial plane of the Kruskal black hole. Jonsson \cite{Jonsson} suggested a way of finding a dual Riemannian metric, geodesically equivalent to $(1+1)$-dimensional static, diagonal Lorentzian metrics. He also proposed a way to visualize curvature in more dimensions, based on finding what he called local Minkowski systems \cite{Jonsson2}.

In this paper we will focus on an idea, introduced by L.\ C.\ Epstein in his book ``Relativity Visualized'' \cite{Epstein}, that, when studying timelike separated events in static spacetimes, we can obtain a Riemannian manifold by measuring distances using the time coordinate $t$, rather than the proper time $\tau$. On this dual Riemannian manifold we are then able to easily visualize the geodesics. An application of this idea to the case of a uniform gravitational field was first explored by Rowland \cite{Rowland}: using what he called an \textit{``Epstein chart''} to study this type of $1$-dimensional motions, he was able to determine the equations of motion without resorting to the standard method -- variational principles and the Euler-Lagrange equations \cite{hartle}. Moreover, he proved that for the case of $1$-dimensional motion the geodesics obtained from the \textit{``Epstein chart''} match those obtained from the Lorentzian manifold.

We aim to explore this idea beyond the \textit{``Epstein chart''} and prove that it provides a useful tool for decoding the properties of static spacetimes. We start by proving that this geodesic correspondence holds for $4$-dimensional manifolds. We explore static spacetimes of constant curvature, and show that they map to constant curvature Riemannian manifolds, whose geodesics are easily visualized. When the resulting Riemannian manifold is the sphere $S^4$, we are able to identify isochronous oscillatory motions. We establish the general conditions required to map to the sphere and thus determine under which circumstances one can find such motions. We look for physically reasonable spacetimes satisfying these conditions in the case of a perfect fluid and of an Einstein cluster. Finally, we study the motions on some non-constant curvature spacetimes by focusing on certain types of motion: the radial motion of massive bodies and the motion of light rays in the equatorial plane. To do so, we present the numerical result obtained for the 2-surfaces corresponding to each case.

We adopt a system of units for which $c=G=1$. We used {\sc Mathematica} for symbolic and numeric computations, and also to produce the figures.

\subsection{Geodesic correspondence}

The metric for a static spacetime can be written in the form
\begin{equation}
    ds^2 = -e^{2\Phi(x^1,\, x^2,\, x^3)} dt^2 + \gamma_{ij}(x^1, x^2, x^3) dx^i dx^j,
    \label{Eq::Geodesics-ST}
\end{equation}
where $\gamma$ stands for a 3-dimensional Riemannian metric. In \cite{Epstein}, L.\ C.\ Epstein suggested that, for timelike separated events, one would get a geodesically equivalent Riemannian metric by rewriting \eqref{Eq::Geodesics-ST} as
\begin{equation}
     dt^2 =e^{-2 \Phi}\left ( d\tau^2 + \gamma_{ij}dx^i dx^j \right )
     \label{Eq::geodesics}
\end{equation}
where $d\tau^2 = -ds^2$ is the proper time interval. Note that $\tau$ is now a coordinate function while $t$ is the arclength.
This establishes a one-to-one correspondence between static spacetimes and Riemannian manifolds, both with the same topology $\mathbb{R} \times \Sigma$, where $\Sigma$ is the $3$-manifold with coordinates $(x^1,x^2,x^3)$.

We now prove the equivalence between the causal geodesics of the Lorentzian metric \eqref{Eq::Geodesics-ST} and the Riemannian metric \eqref{Eq::geodesics} obtained as described above. If we start with a 5-dimensional metric\footnote{This idea is similar to the Eisenhart lift \cite{5dimProof}, and may be considered as an application of the Kaluza-Klein trick without the electromagnetic field; it has been used before in the slightly more general context of stationary spacetimes and Randers metrics in \cite{Caponio1, Caponio2, Caponio3, Caponio4, Caponio5}.}
\begin{equation}
    ds^2 = - e^{2\Phi} dt^2 + \gamma_{ij} dx^i dx^j + d\tau^2
    \label{Eq::5dimmetric}
\end{equation}
and choose the coordinate time $t$ as the parameter, then the $5$-dimensional null geodesics (satisfying $ds^2 = 0$) are the geodesics of the \textit{Fermat metric} \cite{Fermat}
\begin{equation}
     dt^2 = e^{-2\Phi} ( \gamma_{ij} dx^i dx^j + d\tau^2 ) ,
     \label{fermat}
\end{equation}
which is precisely the Epstein metric \eqref{Eq::geodesics}. On the other hand, since the metric \eqref{Eq::5dimmetric} is the Cartesian product of the Lorentzian metric \eqref{Eq::Geodesics-ST} by the trivial metric in $\mathbb{R}$, the projection of the 5-dimensional geodesics (in particular null geodesics) on the submanifolds of constant $\tau$ (parameterized by $(t,x^i)$) are the geodesics of the Lorentizian metric \eqref{Eq::Geodesics-ST}. Therefore, both sets of geodesics coincide (up to reparameterization).

Note that the Fermat metric of the 4-dimensional Lorentzian metric \eqref{Eq::Geodesics-ST} is precisely the metric induced by the Epstein metric \eqref{Eq::geodesics} on the surfaces of constant $\tau$; this is what one should expect, since null geodesics satisfy $ds^2 = 0 \Leftrightarrow d\tau^2 = 0$. The Epstein metric can therefore be seen as a kind of generalization of the Fermat metric that applies to timelike geodesics as well. This is especially interesting in light of the many physical insights that have been obtained by considering the Fermat metric \cite{Abramowicz1, Abramowicz2, Abramowicz3, Karlovini, Sonego1, Sonego2}.

\subsection{Constant Curvature Spacetimes}

\label{section2}

In this section, we apply Epstein's idea to constant curvature spacetimes. One can ask whether the behavior of this correspondence is predictable: by starting with a constant curvature spacetime, will we end up with a constant curvature Riemannian manifold? Will the sign of the curvature $K$ of the spacetime propagate to the Riemannian manifold?

In \cite{Rowland}, Rowland argued that the conditions for a flat, $(1+1)$-dimensional \textit{``Epstein chart''} are not met by any asymptotically flat spacetime, such as the one corresponding to a planet or a star. This is what one would expect if there was a connection between constant curvature spacetimes and constant curvature Riemannian manifolds. The examples considered in this section show that such a connection does seem to exist; nonetheless, there is no discernible pattern regarding the sign of the curvature.

\subsubsection{Minkowski Spacetime}

The metric of the flat Minkowski spacetime is given by
\begin{equation}
    ds^2 = -dt^2 + dx^2 +dy^2 + dz^2 .
    \label{Minkowski}
\end{equation}
Applying Epstein's correspondence to this spacetime leads to the following Riemannian metric:
\begin{equation}
    dt^2 = d\tau^2 + dx^2 +dy^2 + dz^2 .
\end{equation}
This is trivially the 4-dimensional Euclidean space, as might be expected from the fact that causal geodesics in Minkowski spacetime deviate linearly.

\subsubsection{Rindler Spacetime}
Applying Epstein's correspondence to Rindler's spacetime
\begin{equation}
    ds^2 = -z^2 dt^2 +dx^2 +dy^2 + dz^2,
\end{equation}
which is a flat space Lorentzian manifold (a wedge in Minkowski spacetime), leads to the following Riemannian metric:
\begin{equation}
    dt^2 = \frac{1}{z^2}\left(d\tau^2 + dx^2 +dy^2+ dz^2 \right) .
    \label{hyper}
\end{equation}
This is the 4-dimensional hyperbolic space, as might be expected from the fact that the Rindler spatial coordinates of nearby causal geodesics (say two parallel timelike lines in Minkowski spacetime) deviate exponentially due to the differential acceleration of the Rindler static observers. In this example the correspondence turns out to be quite unpredictable: starting with a flat space, $K=0$, the resulting Riemannian metric has negative constant curvature, $K=-1$. 

\subsubsection{de Sitter Spacetime}

The metric of the de Sitter spacetime with cosmological constant $\Lambda>0$ is given by
\begin{equation}
    ds^2 = -\left(1- \frac{\Lambda}{3}r^2 \right)dt^2 +\left(1- \frac{\Lambda}{3}r^2 \right)^{-1}dr^2 + r^2 d\Omega^2 ,
\end{equation}
and has positive constant curvature $K = \frac\Lambda3$. Here $d\Omega^2$ stands for the standard metric on $S^2$, $d\Omega ^2 =  d\theta^2 + \sin^2{\theta} d\phi^2  $. Applying Epstein's correspondence to this spacetime yields
\begin{equation}
    dt^2 = \left(1- \frac{\Lambda}{3}r^2 \right)^{-1}\left[d\tau^2 + \left(1- \frac{\Lambda}{3}r^2 \right)^{-1}dr^2 + r^2 d\Omega^2\right]
    \label{DeSitter-Riemann} .
\end{equation}
This Riemannian metric can be shown to have constant negative curvature $K= -\frac{\Lambda}3$. According to Killing-Hopf theorem \cite{hopf, killing}, it must be the metric of the 4-dimensional hyperbolic space. If one writes the 5-dimensional Minkowski spacetime as the Cartesian product of the Milne $(1+1)$-dimensional universe (the interior of the future light cone of a point in the Minkowski $(1+1)$-dimensional spacetime) and the 3-dimensional Euclidean space,
\begin{equation}
    dt^2 = -d\rho^2 + \rho^2 d\tau^2 + dr^2 + r^2 d\Omega^2 ,
\end{equation}
then one can obtain the 4-dimensional hyperbolic space by considering the spacelike hypersurface
\begin{equation}
    \rho ^2 - r^2 = 1 \Leftrightarrow \begin{cases}
 r = \sinh{u}&\\
 \rho = \cosh{u}&
\end{cases} ,
\end{equation}
which leads to $- d\rho^2 + dr^2 = du^2$, and consequently to
\begin{equation}
    dt^2 = du^2 + \cosh^2{u} \, d\tau^2 + \sinh^2{u} \, d\Omega^2.
\end{equation}
This metric may also be obtained from \eqref{DeSitter-Riemann} by setting $\Lambda =3$ (which is just a choice of units, setting the radius of the cosmological horizon to $1$) and
\begin{equation}
\frac{dr}{1-r^2} = du \Leftrightarrow u = \arctanh{r} \Leftrightarrow \tanh{u} = r,
\end{equation}
confirming that \eqref{DeSitter-Riemann} is indeed the metric of the hyperbolic 4-space of curvature $K=-\frac\Lambda3$. This might be expected from the fact that causal geodesics in de Sitter spacetime deviate exponentially due to the repulsive effect of the positive cosmological constant.

\subsubsection{Flat Anti-de Sitter Spacetime}

The metric of the anti-de Sitter spacetime foliated by flat 2-planes, with cosmological constant $\Lambda <0 $, is given by
\begin{equation}
    ds^2 = \frac{\Lambda}{3} r^2 dt^2 + \left(-\frac{\Lambda}{3}r^2 \right)^{-1} dr^2 +r^2 \left(dx^2 + dy^2 \right) ,
\end{equation}
and has negative constant curvature $K=\frac\Lambda3$. Epstein's correspondence leads to the following Riemannian manifold:
\begin{equation}
    dt^2 = \frac{3}{\Lambda r^2} \left[  - d\tau^2 + \frac{3}{\Lambda r^2} dr^2 - r^2 \left( dx^2 + dy^2 \right) \right] .
\end{equation}
The Riemann tensor of this metric can be computed to be zero. Therefore, according to the Killing-Hopf theorem, this metric must be the 4-dimensional Euclidean space written in some coordinates. In fact, taking $\Lambda = -3$ (by a suitable choice of units) and defining $u = \frac{1}{r}$, one can rewrite the Epstein metric in the following way:
\begin{equation}
    dt^2 = du^2 + u^2 d\tau^2  + dx^2 + dy^2 .
\end{equation}
This is the Cartesian product of the metric for the 2-dimensional Euclidean plane in polar coordinates, where $\tau$ is the angular coordinate, by another 2-dimensional Euclidean plane, which is isometric to the 4-dimensional Euclidean space.

\subsubsection{Hyperbolic Anti-de Sitter Spacetime}

The metric for the anti-de Sitter spacetime foliated by hyperbolic 2-planes, with cosmological constant $\Lambda <0 $, is given by
\begin{equation}
    ds^2 = \left(1+ \frac{\Lambda}{3} r^2\right) dt^2 - \left(1+ \frac{\Lambda}{3} r^2 \right)^{-1} dr^2 + r^2 d\theta^2 + r^2 \sinh^2{\theta} d\phi^2 ,
\end{equation}
and has negative constant curvature $K=\frac\Lambda3$. The resulting Riemannian metric is the following:
\begin{equation}
    dt^2 = - \left(1+ \frac{\Lambda}{3} r^2 \right)^{-1} \left[ - \left(1+ \frac{\Lambda}{3} r^2 \right)^{-1} dr^2  + r^2 \left( d\theta^2 + \sinh^2{\theta} d\phi^2 \right) +d\tau^2 \right] .
    \label{ADSHYPER}
\end{equation}
Computing the Riemann tensor we can see that this metric has constant negative curvature $K=\frac\Lambda3$. Thus, taking again into consideration the Killing-Hopf theorem, it must be the metric of the  4-dimensional hyperbolic space. If one writes the metric of the 5-dimensional Minkowski spacetime as the Cartesian product of the Euclidean 2-plane and the Milne $(2+1)$-dimensional universe (the interior of the future light cone of a point in the Minkowski $(2+1)$-dimensional spacetime),
\begin{equation}
    dt^2 = dr^2 + r^2 d\tau^2 - d\rho ^2  + \rho^2\left( d\theta^2 + \sinh^2{\theta} d\phi^2 \right) ,
\end{equation}
then the 4-dimensional hyperbolic space is given by
\begin{equation}
    \rho ^2 - r^2 = 1 \Leftrightarrow \begin{cases}
 r = \sinh{u}&\\
 \rho = \cosh{u}&
\end{cases} ,
\end{equation}
which yields $dr^2 - d\rho ^2 =du^2$ and thus
\begin{equation}
    dt^2 = du^2 + \sinh^2{u} \, d\tau^2 + \cosh^2{u} \left( d\theta^2 + \sinh^2{\theta} d\phi^2 \right) .
\end{equation}
This is also what follows from \eqref{ADSHYPER} by setting $\Lambda = -3$ and
\begin{equation}
\frac{dr}{1-r^2} = du \Leftrightarrow u = \arctanh{r} \Leftrightarrow \tanh{u} = r,
\end{equation}
confirming that \eqref{ADSHYPER} is indeed the metric of the hyperbolic 4-space of curvature $K=\frac\Lambda3$.

\subsubsection{Spherical Anti-de Sitter Spacetime}\label{Spherical ADS}

The metric of anti-de Sitter spacetime foliated by 2-spheres is given by
\begin{equation}
    ds^2 = -\left(1 - \frac{\Lambda}{3} r^2 \right)dt^2  + \left(1 - \frac{\Lambda}{3} r^2 \right)^{-1} dr^2 +r^2 d\Omega^2 ,
\end{equation}
and has negative constant curvature $K = \frac\Lambda3$. This leads to the following Riemannian metric:
\begin{equation}
    dt^2 = \left(1 - \frac{\Lambda}{3} r^2 \right)^{-1}  \left[ d\tau^2 + \left(1 - \frac{\Lambda}{3} r^2 \right)^{-1} dr^2 + r^2 d\Omega^2 \right] .
    \label{metricADSSPHERIC}
\end{equation}
Computing the Riemann tensor we can see that this metric has positive constant curvature $K=-\frac\Lambda3$, and aso, by the Killing Hopf theorem, it must be the metric of the 4-sphere $S^4$. Indeed, the Euclidean metric of $\mathbb{R}^5$ can be written as
\begin{equation}
    dt^2 = dr^2 + r^2 d\tau ^2 + d\rho^2 + \rho^2 d\Omega^2
\end{equation}
and $S^4$ is simply given by
\begin{equation}
    r^2 + \rho ^2 = 1 \Leftrightarrow \begin{cases}
 r = \cos{u}&\\
 \rho = \sin{u}&
\end{cases}  \quad\left(0\leq u\leq \frac{\pi}{2}\right) .
\end{equation}
This yields $dr^2 + d\rho ^2 = du^2$, and therefore
\begin{equation}
    dt^2 = du^2 + \cos^2{u} \, d\tau^2 + \sin^2{u} \, d\Omega^2 ,
    \label{spheric}
\end{equation}
which is exactly the metric obtained in \eqref{metricADSSPHERIC} by setting $\Lambda = -3$ and
\begin{equation}
\frac{dr}{1+ r^2}= du \Leftrightarrow u = \arctan{r} \Leftrightarrow \tan{u} = r.
\end{equation}
Therefore the geodesics of this spacetime are simply the geodesics of $S^4$, which are great circles. We can then conclude that any free-fall trajectory in the anti-de Sitter spacetime is periodic, making it a \textit{Bertrand spacetime}, as defined in \cite{Perlick}. In this paper it was shown that there are only three parametric families of static, spherically symmetric Bertrand spacetimes, and the anti-de Sitter spacetime indeed belongs to one of these: it can be obtained by setting $K=0$, $D=\frac\Lambda3$, $\beta=2$, $G = -\frac6\Lambda$ and redefining $t' = -\frac\Lambda6 t$ on the type II$_{+}$ family.

\subsection{Radial Isochronous Oscillatory Motion}

The results in Section~\ref{Spherical ADS}, relating the geodesics of the AdS spacetime with the geodesics of $S^4$, show that for geodesics in this spacetime (setting $\Lambda=-3$ for simplicity) $\theta$ and $r$ are periodic in the coordinate time $t$ with period $2\pi$, whereas $\tau$ and $\phi$ increase monotonically by $2\pi$ in the same period. This means that the possible motions in the ADS spacetime are periodic in space and isochronous, that is, they all have the same period as measured by both proper time $\tau$ and coordinate time $t$.

From the Newtonian viewpoint, the radial oscillatory motion of a particle around the centre of a massive ball is a well-studied problem, frequently called the \textit{``gravity train problem''} \cite{kleppner}. When this spherical body is uniformly dense, the gravitational force acting on a particle falling along its diameter is exactly that of a simple harmonic oscillator. Consequently, in the non-relativistic approach, such bodies allow particles to radially oscillate sinusoidally with a period independent of amplitude.

A relativistic approach to the same problem was carried by Parker \cite{RelTrain}. His results showed that for a ball of uniform density, the resulting GR effective potential is a function of the ball's radius $R$, and so is the oscillation period. He found that even for particles released from $r<R$ the existence of the spherically symmetric matter outside the trajectory would affect its period. This led him to conclude that, in the relativistic framework, Newton's spherical shell theorem is not valid.

In this section, we use the results of the previous section to find under which conditions we can obtain radial isochronous oscillatory motion.
To study radial oscillatory motion we must set $d\theta = d\phi = 0$. This reduces $S^4$ to $S^2$ with the usual round metric
\begin{equation}
    dt^2 = du^2 + \cos^2{u} \, d\tau^2 .
    \label{S2}
\end{equation}
Starting with an arbitrary static spherically symmetric spacetime
\begin{equation}
    ds^2 = -e^{2\Phi} dt^2 + e^{2\Lambda} dr^2 +r^2d\Omega^2 ,
    \label{metric_general}
\end{equation}
the correspondent Epstein metric,
\begin{equation}
    dt^2 = e^{-2\Phi} d\tau^2 + e^{2\left(\Lambda - \Phi \right)} dr^2 +e^{-2\Phi} r^2 d\Omega^2 ,
\end{equation}
will reduce to \eqref{S2} if and only if
\begin{equation}
    \begin{cases}
e^{2\left ( \Lambda - \Phi \right )}dr^2  = du^2 \\
\cos^2{u} = e^{-2\Phi}
\end{cases} \Leftrightarrow \begin{cases}
\frac{du}{dr}= e^{\Lambda - \Phi} \\
\cos{u} = e^{-\Phi}
\end{cases}, \quad e^{-\Phi} \in [0,1].
\end{equation}
Differentiating the second equation with respect to $r$ yields
\begin{equation}
    \sin{u} \, u' = \Phi' e^{-\Phi} \Leftrightarrow   e^{\Lambda}\sqrt{1 - e^{-2\Phi}} =  \Phi ' \Leftrightarrow e^{\Lambda} = \frac{\Phi ' e^{\Phi}}{\sqrt{ e^{2\Phi} - 1}} ,
    \label{radial_osc}
\end{equation}
where $^\prime$ stands for $\frac{d}{dr}$, giving us the relation between $e^\Lambda$ and $e^\Phi$.
When implementing this condition, it is useful to change of coordinates: defining
\begin{equation}
e^{\Phi} = \cosh{\psi} ,
\label{cosh}
\end{equation}
equation \eqref{radial_osc} leads to
\begin{equation}
    \psi' = e^{\Lambda} \Rightarrow \psi = \int_{0}^{r} e^{\Lambda(s)} ds .
    \label{other}
\end{equation}

For the spherically symmetric metric \eqref{metric_general}, the Einstein tensor reduces to \cite{Schutz}
\begin{align}
    &G_{tt} = \frac{e^{2\Phi}}{r^2} \left[ 1 + e^{-2\Lambda} \left(2r\Lambda^\prime -1 \right)\right],\\
    &G_{rr} = \frac{1}{r^2} \left(1 - e^{2 \Lambda}\right) + \frac{2\Phi^\prime}{r} ,\\
    & G_{\theta \theta} = r^2 e^{-2\Lambda} \left( \Phi^{\prime \prime} + {\Phi}^{\prime 2} + \frac{\Phi^\prime}{r} - \Phi^\prime \Lambda^\prime - \frac{\Lambda^\prime}{r} \right), \\
    & G_{\phi \phi} = \sin^2{\theta} G_{\theta \theta},
\end{align}
with all other components vanishing. For a general diagonal stress-energy tensor, given by
\begin{align}
    &T_{tt} = \rho e^{2\Phi} , \\
    & T_{rr} = p_r e^{2\Lambda} , \\
    & T_{\theta \theta} = r^2 p_{\theta} , \\
    & T_{\phi \phi} = \sin^2{\theta} T_{\theta \theta},
\end{align}
where $\rho$ stands for the energy density, $p_r$ is the radial pressure and $p_{\theta}$ the tangential pressure,
Einstein's equations are given by
\begin{align}
    & \frac{1}{r^2} \left[ 1  + \frac{1}{\psi'} \left( \frac{2r\psi''}{\psi'} - 1\right)\right] = 8 \pi \rho , \\
    & \frac{1-\psi'^2 }{r^2} + \frac{2 \psi' \tanh{\psi}}{r} = 8\pi p_r \psi'^2 , \label{1}\\
    & \frac{1}{\psi'^2} \left(\psi'^2 + \frac{\psi' \tanh{\psi}}{r} - \frac{\psi''}{\psi' r}\right) = 8 \pi p_{\theta}. \label{2}
\end{align}

These equations give the energy density, radial pressure and tangential pressure of the matter generating a spacetime in which there are radial isochronous oscillatory motions. Notice that the choice of $\psi(r)$ is arbitrary, and determines the metric through \eqref{cosh} and \eqref{other}. \par

\subsubsection{Perfect Fluid}
For the case of a perfect fluid, we must have $p_r = p_\theta$, and so by \eqref{1} and \eqref{2}, $\psi$ must satisfy
\begin{equation} \label{equation48}
    - \psi^{\prime 3} + \psi^\prime + r \psi^{\prime 2} \tanh{\psi} - r^2 \psi^{\prime 3} + \psi^{\prime \prime} r = 0 .
\end{equation}
Numerically solving this equation leads to various solutions for $\psi(r)$, $\psi'(r)$ and, consequently, for $e^{\Lambda(r)}$. However, only a one-parameter family of these solutions satisfy the condition that both the radial pressure and the density remain finite at the center $r=0$. Below we present one of these solutions and the corresponding $p_r (r)$ and $\rho (r)$ profiles. The initial values for the numerical solution were given at $r= 0.75$, and $r$ ranges from $r= 0.001$ to $r=0.75$.

\begin{figure}[h!]
\begin{subfigure}{.5\linewidth}
\centering
\includegraphics[scale=0.18]{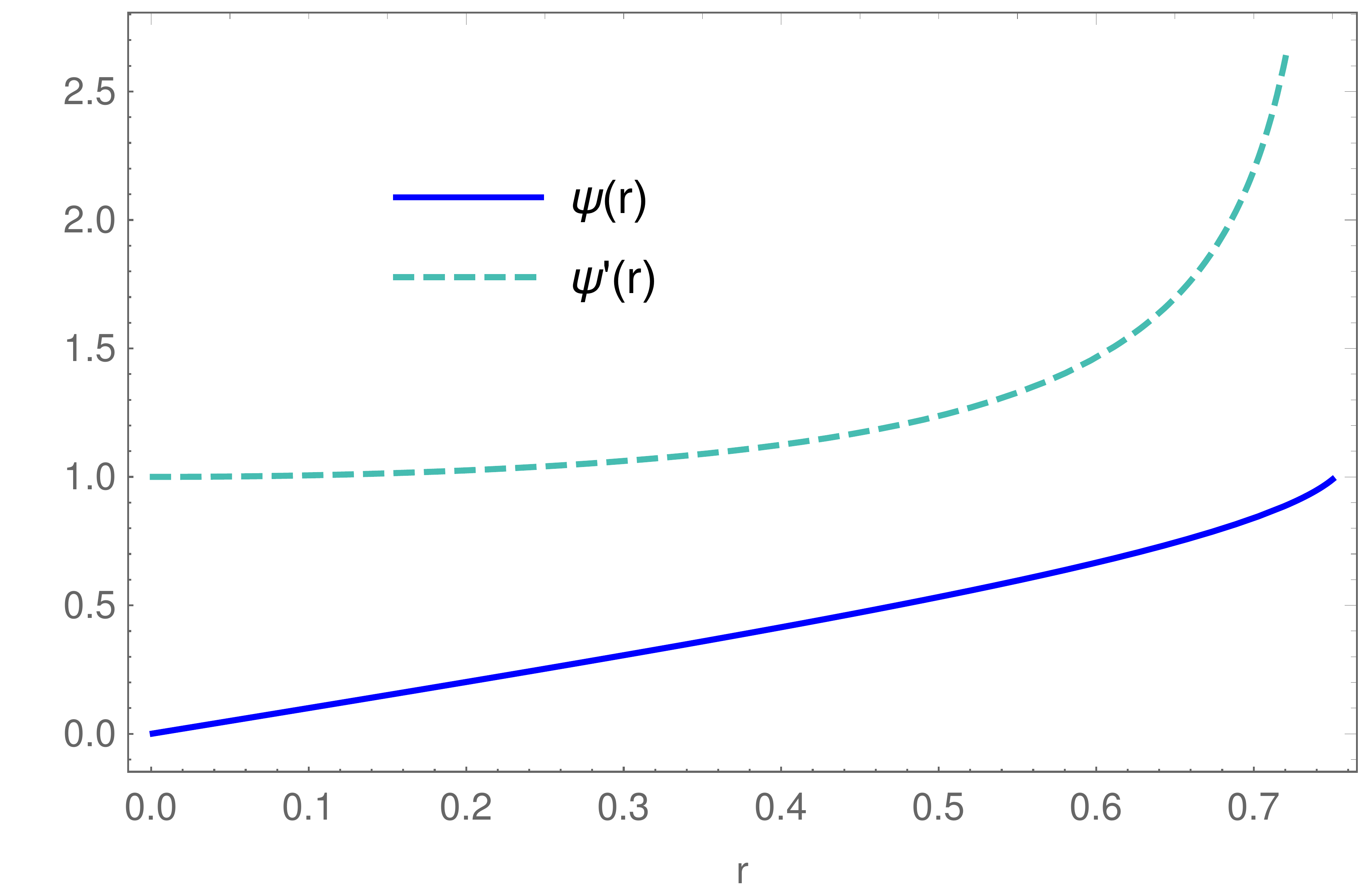}
\end{subfigure}
\begin{subfigure}{.5\linewidth}
\centering
\includegraphics[scale=0.18]{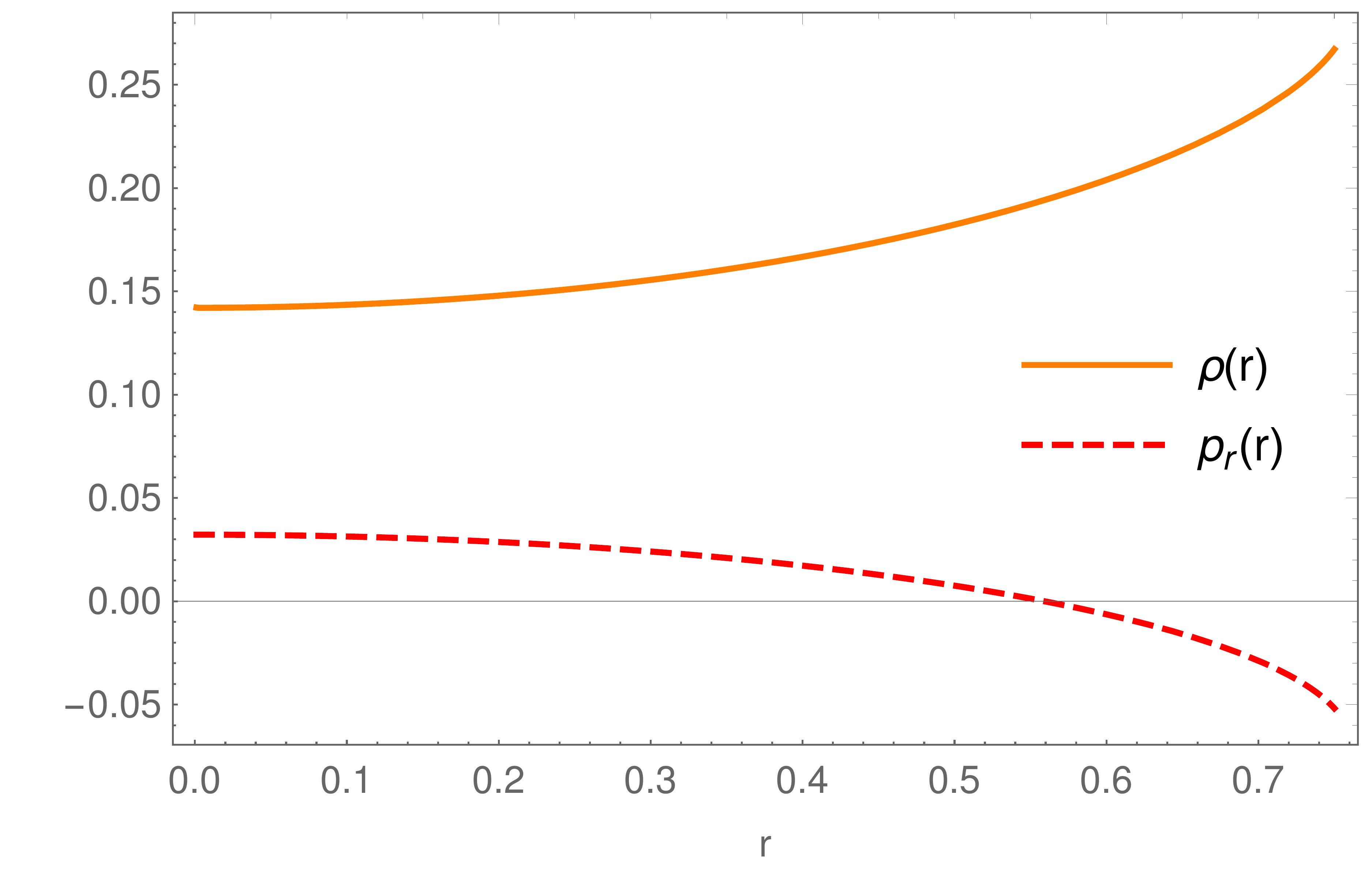}
\end{subfigure}
\caption{Left: solution of equation~\eqref{equation48} corresponding to $\psi(0.75)=0.99$ and $\psi'(0.75)=4.9915$. Right: corresponding pressure and density profiles.}
\end{figure}

Looking at the second plot, one can check that the solution satisfy the dominant energy condition, $\rho> \left | p_r \right |$, and that $p_r$ vanishes at a certain value of $r$, as one would expect for a finite body. However, the second plot shows that the radial pressure decreases as the density grows. This means that the equation of state, $p=p(\rho)$, will be such that $\frac{dp}{d\rho}<0$ and, consequently, the fluid is unstable against small perturbations. This behavior was found in all the numerical solutions obtained, and so it appears that only unstable fluids may give rise to isochronous oscillatory motion.

\subsubsection
{Einstein Cluster} \textit{``Einstein cluster''} refers to a class of solutions of Einstein's equations proposed by Einstein in 1939 \cite{Einstein}. It models a cloud of massive particles following circular geodesics in all directions around a common center while being acted on by their collective gravitational field. In this system, the radial pressure $p_r$ vanishes, and so equation \eqref{1} reduces to
\begin{equation} \label{equation49}
    \frac{1-\psi'^2 }{r^2} + \frac{2 \psi' \tanh{\psi}}{r} = 0 .
\end{equation}
Numerically solving this differential equation leads to a single physically meaningful solution (in which all physical quantities remain finite) for each radius chosen to give the initial condition. In the plot below, we present the solution where the initial condition is given at $r=0.5$, and the corresponding $p_{\theta}(r)$ and $\rho(r)$ profiles.

\begin{figure}[h!]
\begin{subfigure}{.5\linewidth}
\centering
\includegraphics[scale=0.18]{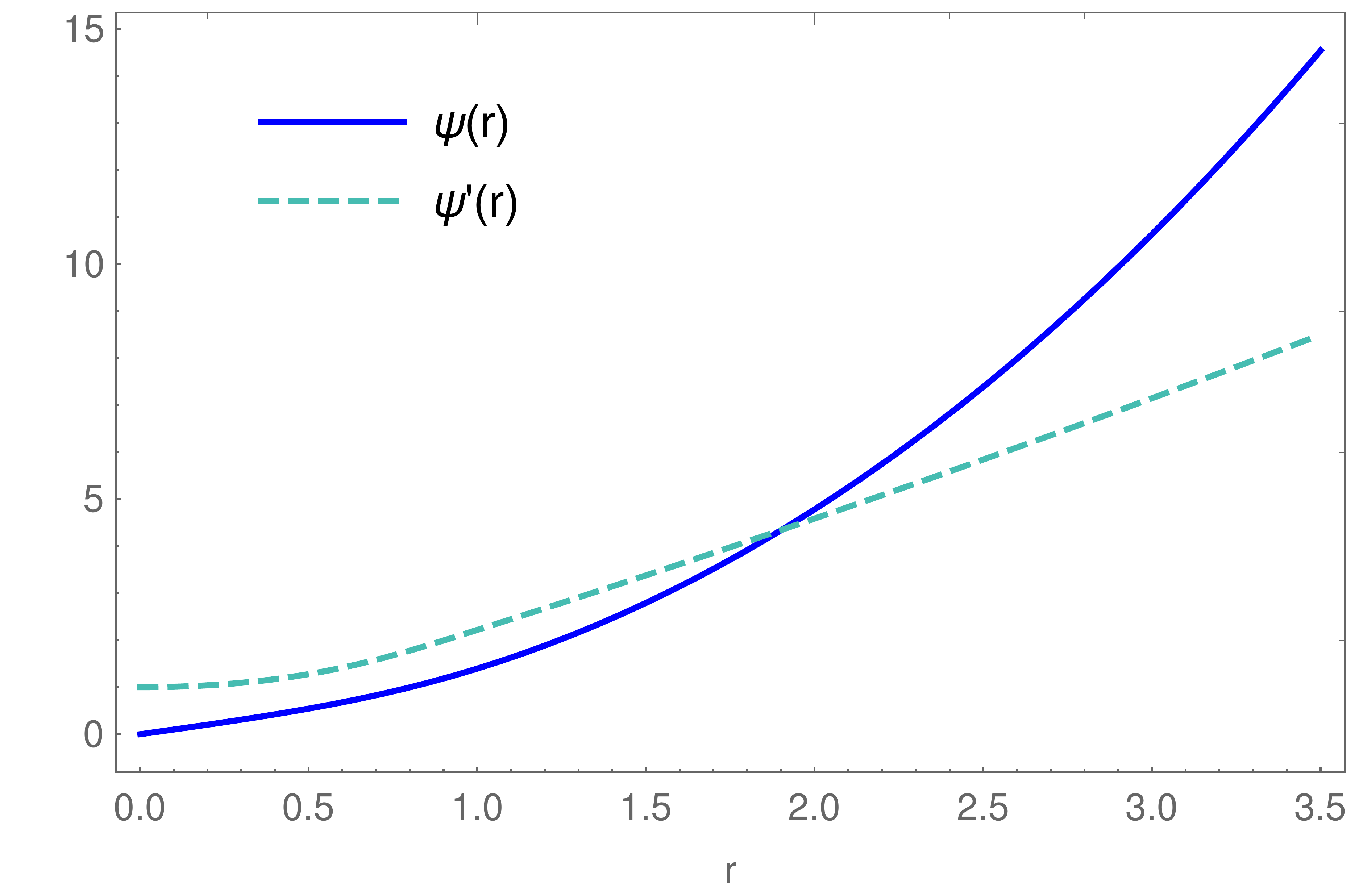}
\end{subfigure}
\begin{subfigure}{.5\linewidth}
\centering
\includegraphics[scale=0.18]{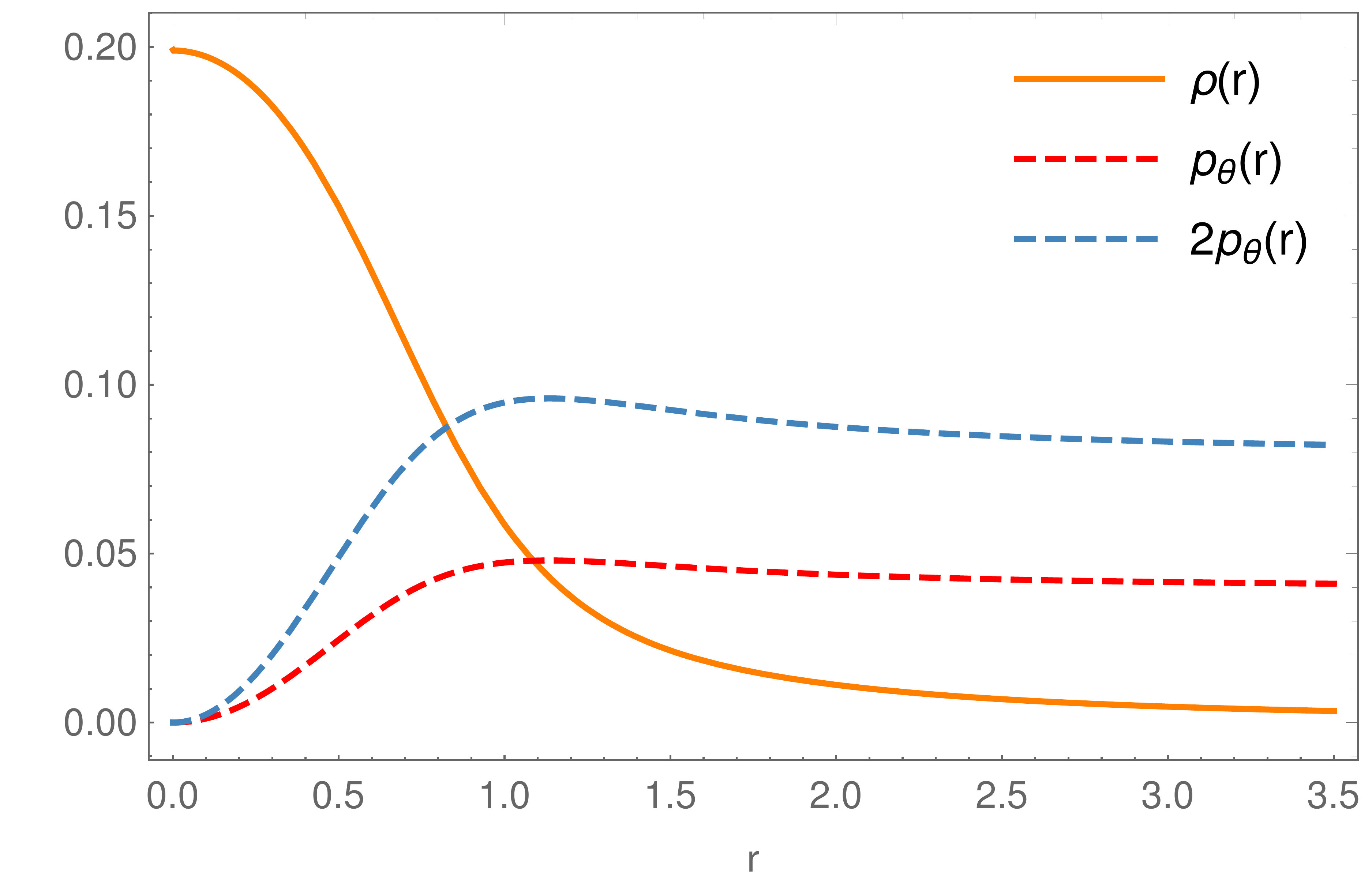}
\end{subfigure}
\caption{Left: solution of equation~\eqref{equation49} corresponding to $\psi(0.5)=0.54462$. Right: corresponding tangential pressure and density profiles.}
\end{figure}

These solutions describe an Einstein cluster of particles moving slower than the speed of light as long the condition $\rho> 2 p_\theta > 0$ is satisfied. Therefore, the results obtained allow us to conclude that there are Einstein clusters which allow for radial isochronous oscillatory motion.

\subsection{Non-Constant Curvature Spacetimes}

To model the gravitational field of planets, stars or black holes we must use non-constant curvature spacetimes. The Riemannian 4-manifolds obtained using Epstein's correspondence in these situations do not have constant curvature either, and so we cannot easily identify their geodesics. For simplicity, we restrict our study to two different types of motion: the radial motion of massive particles and the motion of light rays in the equatorial plane. By doing so, we lower the dimension of the Riemannian manifolds to $2$, allowing us to embed them into the 3-dimensional Euclidean space as surfaces of revolution, where the task of decoding the geodesics is much simpler.

\subsubsection{Schwarzschild Spacetime} \label{Schwarzschildsection}

The Schwarzschild metric describes the gravitational field produced by a spherical body of mass $m$:
\begin{equation}
  ds^2 = -\left( 1 - \frac{2m}{r}\right) dt^2 +\left( 1 - \frac{2m}{r}\right)^{-1} dr^2 + r^2 d\Omega^2 .
  \label{SCHWARZ}
\end{equation}
This metric has vanishing Ricci tensor (and consequently vanishing scalar curvature). Taking into account the results of Section~\ref{section2}, one might expect
that the resulting Epstein metric,
\begin{equation}
  dt^2 = \left( 1 - \frac{2m}{r}\right)^{-1} \left[ d\tau^2  +  \left( 1 - \frac{2m}{r}\right)^{-1} dr^2 + r^2 d\Omega^2 \right],
  \label{EPSSCHWARZ}
\end{equation}
would also have vanishing Ricci tensor. However, this is not the case, since it has scalar curvature $S = -\frac{12 m^2}{r^4}$. We will now study some of its
totally geodesic surfaces associated to certain types of motion.

\subsubsection*{ Null Geodesics }
For the motion of a light ray in the equatorial plane, for example, we have $d\tau^2 = 0$ and $\theta = \frac{\pi}{2} \Rightarrow d\theta^2 = 0$, and so the Epstein metric
reduces to
\begin{equation}
  dt^2 = \left(1-\frac{2m}{r}\right)^{-1}\left[ \left(1-\frac{2m}{r}\right)^{-1} dr^2 + r^2 d\phi^2 \right].
\end{equation}

\begin{figure}[h!]
    \centering
    \includegraphics[scale = 0.6]{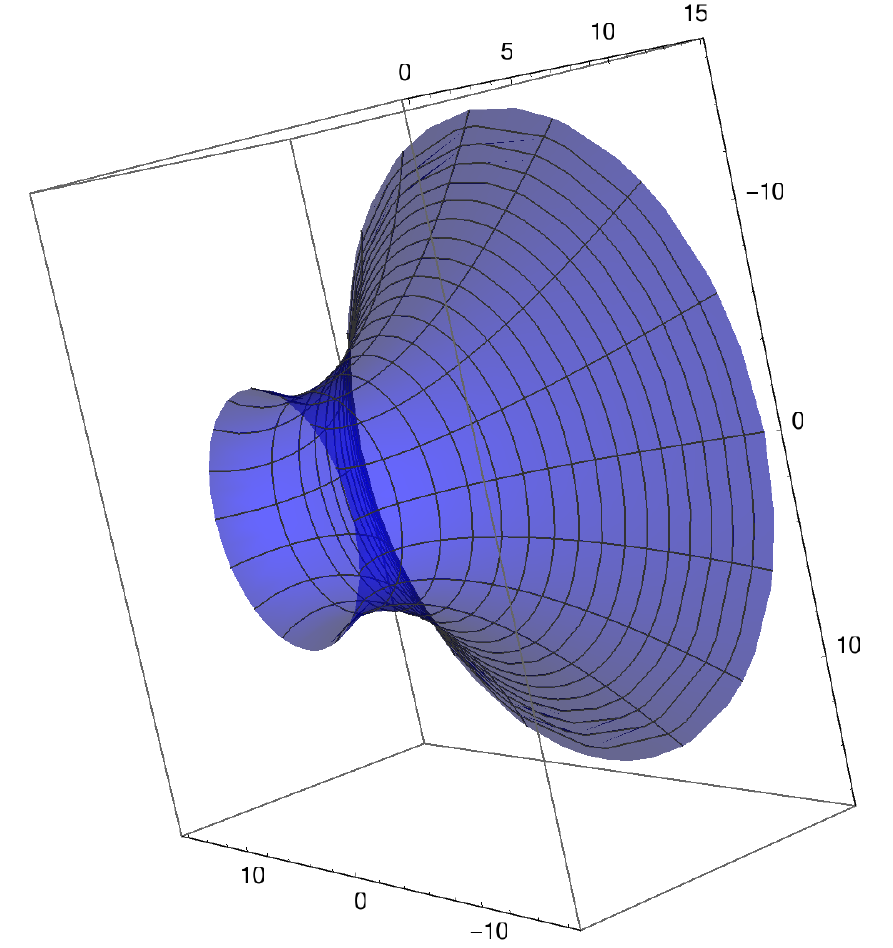}
    \caption{Epstein surface for the motion of light rays in the equatorial plane of the Schwarschild spacetime ($d\tau = d\theta =0$).}
    \label{fig:Schwarz}
\end{figure}

In order to visualize the resulting manifold we use cylindrical coordinates $(\rho,\phi,z)$ and set $\rho^2 = \left(1-\frac{2m}{r}\right)^{-1} r^2$ and $d\rho^2 + dz^2 =  \left(1-\frac{2m}{r}\right)^{-2} dr^2$. From the second condition, one can find $\frac{dz}{dr}$ and therefore extract $z(r)$. The result obtained for unit mass, $m=1$, is presented in Figure~\ref{fig:Schwarz}.

The resulting manifold is a surface of revolution, given by $ \rho = f(z)$ for some function $f$, and consequently, $ds^2 = f(z)^2 d\phi^2 + \left(f'^2(z) + 1\right)  dz^2$. The geodesic Lagrangian for this surface is $L = \frac{1}{2}\left(f^2(z) \dot{\phi}^2 + \left(f'^2(z) + 1\right) \dot{z}^2 \right)$. Since $\phi$ is a cyclic coordinate, one can easily identify the effective potential of the system: $ U_{eff} = \frac{l_{\phi}}{f^2(z)} $, with $l_{\phi}$ the conserved momenta in $\phi$. The curve in the plot determined by the minimum of $ \rho = f(z)$ corresponds to a maximum of $U_{eff}$ and, thus, to an unstable circular geodesic, located at $\frac{d\rho}{dr} = 0 \Leftrightarrow r= 3m$ (corresponding to the well known photonsphere; this geometrical feature has been used to interpret phenomena such as the existence of trapped modes of gravitational waves or the reversal of the centrifugal force for circular trajectories around a Schwarzschild black hole \cite{Abramowicz1, Abramowicz2, Abramowicz3}). Note that any small perturbation leads to motion either towards infinity or towards the black hole. Moreover, since lengths on this manifold correspond to measurements of coordinate time, we can conclude that this orbit yields the fastest way that any particle can circle a black hole, in agreement with \cite{FastestWayCircleBH}. Other obvious geodesics are the meridians of the surface, corresponding to radial light rays.

\subsubsection*{Radial Motion}
\label{radial motion}
\begin{figure}[h!]
    \centering
    \includegraphics{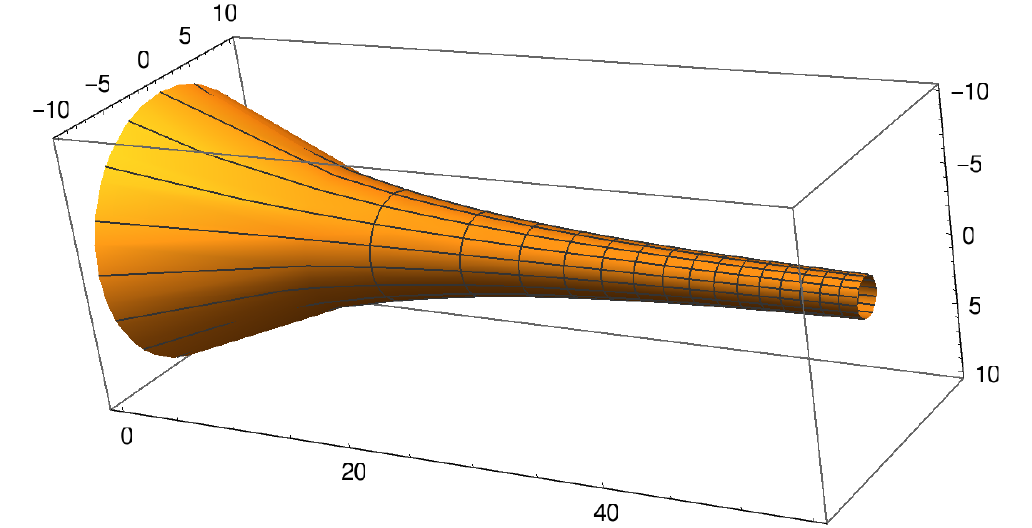}
    \caption{Epstein surface for radial motion in the Schwarschild spacetime ($d\theta = d\phi =0$).}
    \label{fig:Schwarzdphi}
\end{figure}

We can also visualize the surface describing radial motion ($d\theta = d\phi = 0$)  as a surface of revolution by taking $\tau$ to be an angular coordinate: defining cylindrical coordinates $(\rho, \tau,z)$ and setting $\rho^2 = \left(1-\frac{2m}{r}\right)^{-1}$ and $d\rho^2 + dz^2 =  \left(1-\frac{2m}{r}\right)^{-2} dr^2$, we obtain the surface depicted in Figure~\ref{fig:Schwarzdphi}.

Since there are no maxima or minima of $ \rho = f(z)$, there are no geodesics of constant $r$, and the possible trajectories for free-falling particles correspond to geodesics that wind around the surface. When $t \rightarrow \pm \infty$, they satisfy either $\rho \to \infty$, which corresponds to the black hole event horizon, or $\rho \to 1$, which corresponds to spatial infinity. Trajectories of light rays correspond to meridians, since $d\tau^2=0$, and so do not wind around the surface but instead go straight from the event horizon to infinity or vice-versa (the black lines in the plot). Note that far from the black hole the surface resembles a flat cylinder, as might be expected, since it must approach the Minkowski spacetime, whose Epstein dual is also flat.

\subsubsection{Interior Solution}

A possible interior solution for the Schwarschild spacetime is obtained by choosing constant matter density, and is given by the metric
\begin{equation}
    ds^2 = -\left[ \frac{3}{2} \left( 1 - \frac{2m}{R}\right)^{\frac{1}{2}} - \frac{1}{2} \left(1- \frac{2m}{R^3} r^2 \right)^{\frac{1}{2}} \right]^{2} dt^2 +  \left(1- \frac{2m}{R^3} r^2 \right) dr^2 + r^2 d\Omega^2,
\end{equation}
where $R$ stands for the radius of the spherical body. The resulting Epstein metric is given by
\begin{equation}
    dt^2 = \left[ \frac{3}{2} \left( 1 - \frac{2m}{R}\right)^{\frac{1}{2}} - \frac{1}{2} \left(1- \frac{2m}{R^3} r^2 \right)^{\frac{1}{2}} \right]^{-2} \left[ d\tau^2 + \left(1- \frac{2m}{R^3} r^2 \right) dr^2 + r^2  d\Omega^2 \right].
\end{equation}
These metrics are valid for $r<R$, and are continuously extended by \eqref{SCHWARZ} and \eqref{EPSSCHWARZ} for $r>R$.

\subsubsection*{Null geodesics}

\begin{figure}[h!]
    \centering
    \includegraphics[scale =0.8]{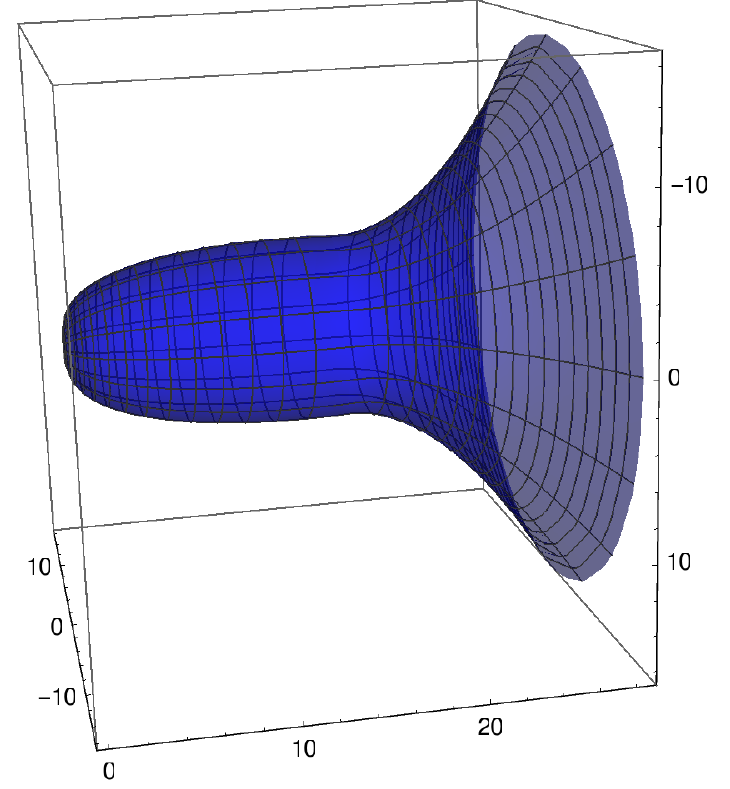}
    \caption{Epstein surface for the motion of light rays in the equatorial plane of the Schwarschild spacetime with a constant density interior ($d\tau = d\theta =0$).}
    \label{SchwarzIntdTau}
\end{figure}

For the motion of a light ray in the equatorial plane, one must set $d\theta = d\tau =0$ and choose $\rho$ and $z$ so that, for $r<R$,
\begin{equation}
\rho^2 = \left[ \frac{3}{2} \left( 1 - \frac{2m}{R}\right)^{\frac{1}{2}} - \frac{1}{2} \left(1- \frac{2m}{R^3} r^2 \right)^{\frac{1}{2}} \right]^{-2} r^2
\end{equation}
and
\begin{equation}
d\rho^2 + dz^2 = \left[ \frac{3}{2} \left( 1 - \frac{2m}{R}\right)^{\frac{1}{2}} - \frac{1}{2} \left(1- \frac{2m}{R^3} r^2 \right)^{\frac{1}{2}} \right]^{-2} \left(1- \frac{2m}{R^3} r^2 \right) dr^2.
\end{equation}
The surface obtained setting $R = 2.8$ and $m=1$ is presented in Figure~\ref{SchwarzIntdTau}.

The meridians of this surface (the black lines in the plot) correspond to the motion of light rays with no angular momenta, $d\phi =0$, that simply go through our star.
From the shape of this surface we may conclude that the coordinate time necessary to cross the star, that is, the length of the geodesic, is larger than it would be in a flat space. This corresponds to well known Shapiro effect \cite{PhysRevLett.13.789}.
In the case depicted in Figure~\ref{SchwarzIntdTau}, one can still see the minimum of $\rho=f(z)$ previously identified at $r=3m$, which matches an unstable circular geodesic. Now, however, small perturbations towards the center will lead to geodesics that winds around the surface towards $r=0$ and then returns back to $r=3m$, possibly taking infinite time to complete this cycle. Note that there is also a maximum of $\rho=f(z)$ inside the constant density region, corresponding to a stable circular orbit, in agreement with \cite{Herdeiro}.

\subsubsection*{Radial Motion}

\begin{figure}[h!]
    \centering
    \includegraphics[scale=0.5]{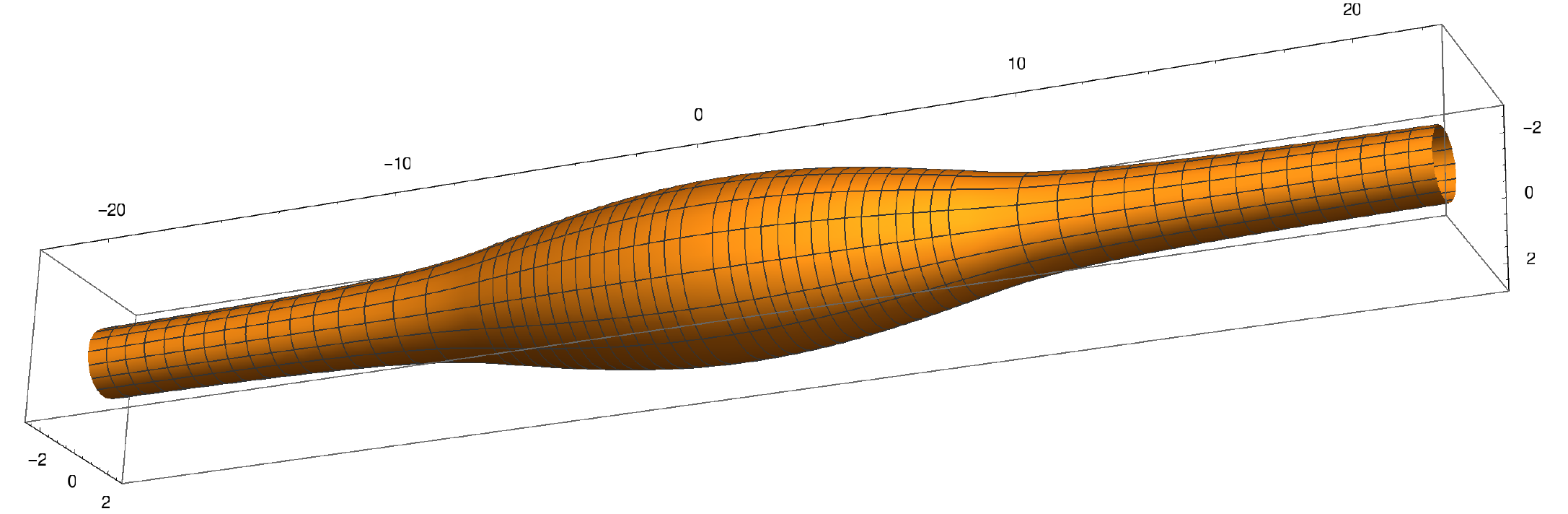}
    \caption{Epstein surface for radial motion in the Schwarschild spacetime with a constant density interior ($d\theta = d\phi =0$).}
    \label{SchwarzIntDtheta}
\end{figure}

Setting $d\theta= d\phi=0$ and defining $\rho$ and $z$ such that, for $r<R$,
\begin{equation}
\rho^2 =  \left[ \frac{3}{2} \left( 1 - \frac{2m}{R}\right)^{\frac{1}{2}} - \frac{1}{2} \left(1- \frac{2m}{R^3} r^2 \right)^{\frac{1}{2}} \right]^{-2}
\end{equation}
and
\begin{equation}
d\rho^2 + dz^2 =  \left[ \frac{3}{2} \left( 1 - \frac{2m}{R}\right)^{\frac{1}{2}} - \frac{1}{2} \left(1- \frac{2m}{R^3} r^2 \right)^{\frac{1}{2}} \right]^{-2} \left(1- \frac{2m}{R^3} r^2 \right) dr^2,
\end{equation}
one obtain the surface describing radial motion. The result, choosing $R = 3$ and $m=1$, is shown in Figure~\ref{SchwarzIntDtheta} (where points with $z<0$ represent points antipodal to those with $z>0$).

In addition to the geodesics identified in Section~\ref{radial motion}, the interior solution introduces new possibilities of motion: for instance, $\rho = f(z)$ reaches a maximum for $z=0$. Defining the Lagrangian $L = \frac{1}{2} \left( f^2(z) \dot{\tau}^2 + \left(f'(z)^2 + 1 \right)\dot{z}^2 \right)$ as before, we can conclude that the effective potential $U_{eff} = \frac{l_{\tau}}{f^2(z)}$ has a minimum at $z=0$. Consequently, there is a stable circular geodesic at $z=0$. This represents the situation of a particle at rest in the centre of our star. Any small perturbation leads to oscillatory motion represented by geodesics winding around the surface around $z=0$. This type of motion corresponds to radial oscillations near the centre of our massive body. Null geodesics, the motion of light rays, correspond to the black lines in the plot. Since $d\tau = 0$, there is no angular motion around the surface and, therefore, light rays simply go through our massive body.
\par

\subsubsection{Schwarzschild de Sitter Spacetime}

The Schwarzschild de Sitter spacetime with cosmological constant $\Lambda > 0$ is given by the metric
\begin{equation}
    ds^2 = - \left( 1 - \frac{2m}{r} - \frac{\Lambda}{3}r^2 \right) dt^2 + \left( 1 - \frac{2m}{r} - \frac{\Lambda}{3}r^2 \right)^{-1} dr^2 + r^2d\Omega^2 ,
\end{equation}
and contains both a black hole event horizon and a cosmological horizon. The corresponding Epstein metric is given by
\begin{equation}
    dt^2 = \left( 1 - \frac{2m}{r} - \frac{\Lambda}{3}r^2 \right)^{-1} \left[ d\tau ^2 + \left( 1 - \frac{2m}{r} - \frac{\Lambda}{3}r^2 \right)^{-1}  dr^2 + r^2 d\Omega^2 \right] .
\end{equation}
Again, this metric does not have even constant scalar curvature, despite of the fact that its Lorentzian dual is an Einstein manifold.

\subsubsection*{Radial Motion}

\begin{figure}[h!]
    \centering
    \includegraphics[scale = 0.7]{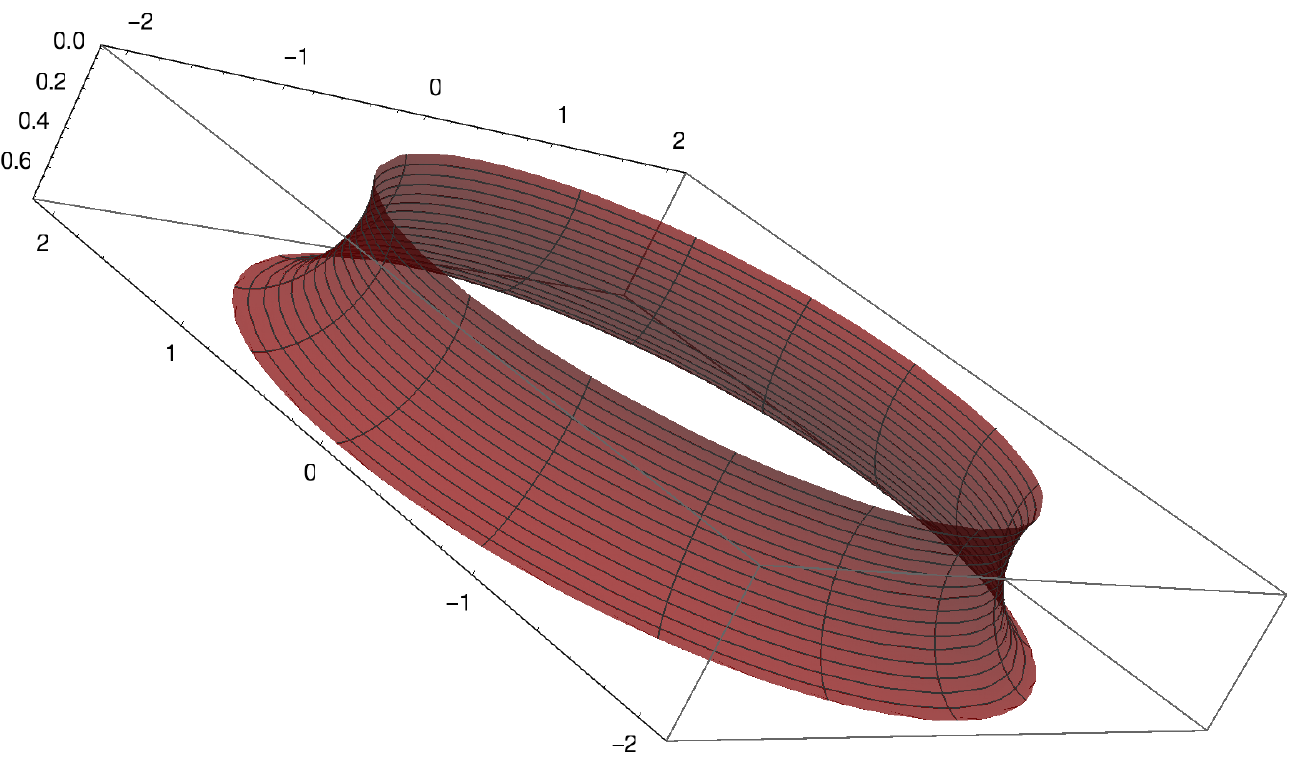}
    \caption{Epstein surface for radial motion in the Schwarschild de Sitter spacetime ($d\theta = d\phi =0$).}
    \label{SCHWARZSCHILDDESITTERDPHI0}
\end{figure}

Setting $d\theta = d\phi = 0$ and defining $\rho$ and $z$ such that $\rho^2 =  \left( 1 - \frac{2m}{r} - \frac{\Lambda}{3}r^2 \right)^{-1} $ and $d\rho^2 + dz^2 =  \left( 1 - \frac{2m}{r} - \frac{\Lambda}{3}r^2 \right)^{-2} dr^2$, for $m= 0.1$ and $\Lambda = 3$, we obtain the surface presented in Figure~\ref{SCHWARZSCHILDDESITTERDPHI0}.

From the plot one can easily see that $\rho=f(z)$ has a minimum, which signals an unstable circular geodesic at $\frac{d\rho}{dr} = 0 \Leftrightarrow r= \left(\frac{3m}{\Lambda}\right)^{\frac{1}{3}}$, corresponding to an equilibrium position for massive particles. This is due to the repulsive character of the positive cosmological constant: at $r= \left(\frac{3m}{\Lambda}\right)^{\frac{1}{3}}$, the repulsive cosmological force balances the attractive gravitational force created by the black hole. Any small perturbation leads either to motion towards the black hole horizon or towards the cosmological horizon, given by geodesics that wind around the surface. For light rays, $d\tau = 0$, the geodesics are again the meridians of our surface, represented by the black lines in the plot, going either towards the black hole horizon or the cosmological horizon.

\subsubsection{Schwarzschild Anti-de Sitter Spacetime}

The Schwarzschild anti-de Sitter spacetime is given by the exact same metric as the Schwarzschild de Sitter spacetime, except for the fact that now $\Lambda < 0$, and consequently there is no cosmological horizon.

\subsubsection*{Radial Motion}
\begin{figure}[h!]
    \centering
    \includegraphics{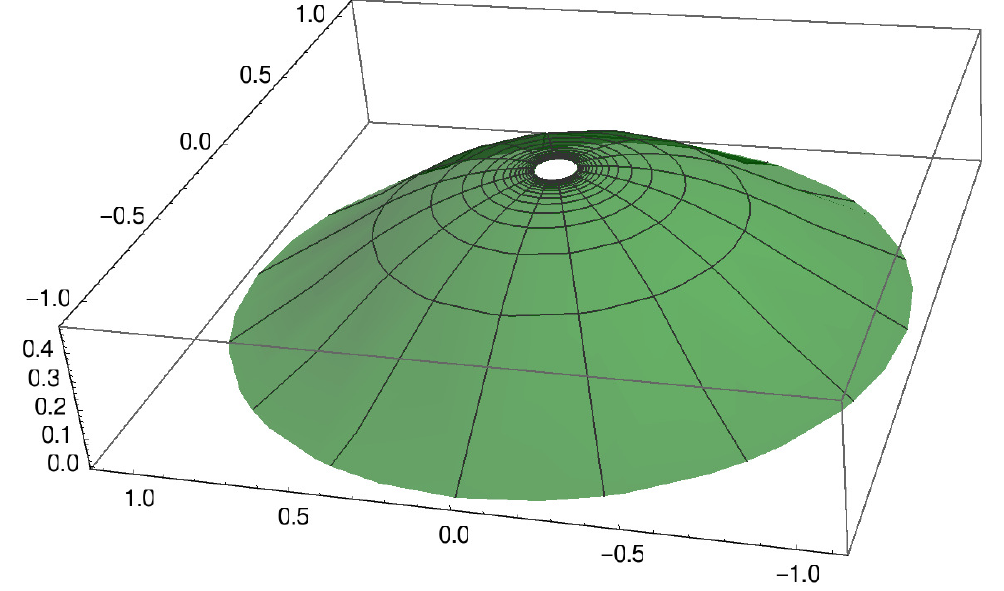}
    \caption{Epstein surface for radial motion in the Schwarschild anti-de Sitter spacetime ($d\theta = d\phi =0$).}
    \label{SCDSPHI0}
\end{figure}

Following the same procedure as before, for $m = 0.1$ and $\Lambda = -3$ we obtain the surface depicted in Figure~\ref{SCDSPHI0}, where the ``hole'' on the top is actually just a missing point corresponding to infinity (since the Schwarzschild anti-de Sitter metric approaches the anti-de Sitter metric at infinity, its Epstein metric approaches the metric of the sphere).

For a negative cosmological constant there is no sign of unstable circular geodesics. Light rays correspond to geodesics given by the black lines, $d\tau = 0$, and so they travel either towards the black hole or towards infinity. By opposition, particles moving slower that light may move towards infinity initially, but will eventually return to $r=0$. This is due to the attractive character of the Schwarzschild Anti-de Sitter spacetime, as in this case both the black hole and the cosmological constant attract the particles towards $r=0$.

\subsubsection{Wormholes}

According to \cite{Wormholes}, the spacetime metric that describes a static and spherically symmetric
wormhole has the form
\begin{equation}
    ds^2 = - e ^{2\Phi(r)} dt^2 + \frac{dr^2}{1 - \frac{b(r)}{r}} + r^2 d\Omega^2 ,
\end{equation}
where $\Phi(r)$ and $b(r)$ are arbitrary functions of the radial coordinate $r$.
While $\Phi(r)$ describes the gravitational redshift, $b(r)$ determines the shape of the wormhole. The radial coordinate ranges from
$r_0$, the wormhole's throat, to $a$, the wormhole's mouth. At  $r=r_0$ one should mirror this spherical volume to a copy such that $r$ goes again
from $r_0$ to $a$. Additionally, one must join to each copy the desired external spacetime with $r$ going from $a$ to $\infty$, ensuring continuity at $r=a$.
This Lorentzian metric leads to the following Epstein metric:
\begin{equation}
    dt^2 = e ^{-2\Phi}\left( d\tau^2 + \frac{dr^2}{1 - \frac{b(r)}{r}} + r^2 d\Omega^2\right) .
\end{equation}
We will now explore an example presented in \cite{Wormholes}, where a matching of an interior solution
to an exterior Schwarzschild solution was considered with zero tangential pressure at the junction. Under these circumstances, if follows from \cite{Wormholes} that for a wormhole of unit mass, $m = 1$, we have $b(a) = 2$. Choosing
\begin{align}
  & \Phi(r) = \Phi_0 ,\\
  & b(r) = (r_0 r) ^{\frac{1}{2}},
\end{align}
we have $b(a) = (r_0 a)^{\frac{1}{2}}$, and so the matching happens at $a = \frac{4}{r_0}$.
Imposing continuity at $r=a $, the interior metric ($r_0 \leq r \leq a$) is given by
\begin{equation}
  ds^2 = - \left( 1 - \sqrt{\frac{r_0}{a}}\right)dt^2 + \frac{dr^2}{\left( 1 - \sqrt{\frac{r_0}{r}}\right)} + r^2 d\Omega^2 ,
\end{equation}
and the exterior metric ($r \geq a$) is given by
\begin{equation}
  ds^2 = - \left( 1 - \frac{\sqrt{r_0 a}}{r}\right)dt^2 + \frac{dr^2}{\left( 1 - \frac{\sqrt{r_0 a}}{r}\right)} + r^2 d\Omega^2 ,
\end{equation}
which leads us to the following Epstein metric:
\begin{equation}
\begin{cases}
dt^2 = \left ( 1 - \sqrt{\frac{r_0}{a}} \right )^{-1} \left [ d\tau^2 + \left ( 1 - \sqrt{\frac{r_0}{r}} \right )^{-1}dr^2 + r^2 d\Omega^2 \right ] & \text{ ,  } \quad r_0\leq r\leq a \\
dt^2 = \left ( 1 - \frac{\sqrt{r_0 a}}{r} \right )^{-1} \left [ d\tau^2 + \left ( 1 - \frac{\sqrt{r_0 a}}{r} \right )^{-1}dr^2 + r^2 d\Omega^2 \right ] & \text{ , } \quad  r\geq a
\end{cases} .
\end{equation}

\subsubsection{Radial Motion}

\begin{figure}[h!]
  \centering
  \includegraphics[scale=0.8]{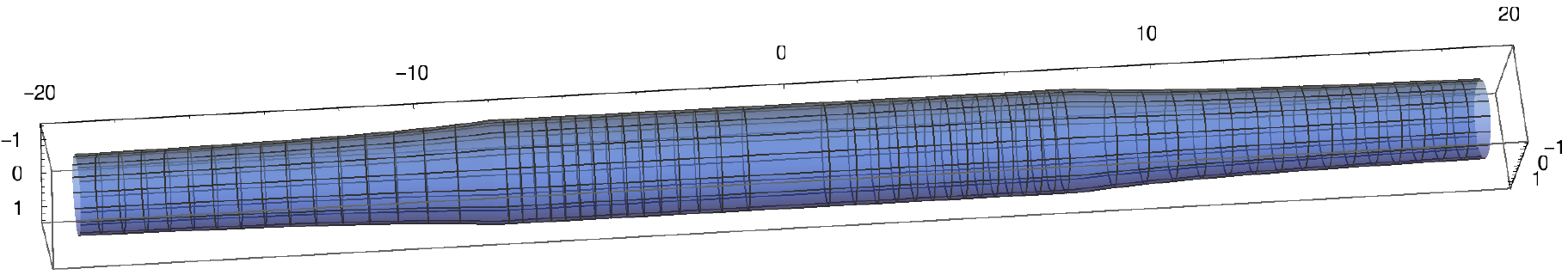}
  \caption{Epstein surface for radial motion in the wormhole spacetime ($d\theta = d\phi =0$).}
  \label{fig:wormhole}
\end{figure}

We focus on radial motion by setting $d\theta = d\phi = 0$ and defining $\rho$ and $z$ such that:
\begin{equation}
\begin{cases}
\rho^2 =   \left ( 1 - \sqrt{\frac{r_0}{a}} \right )^{-1} \text{ and } d\rho^2 + dz^2 = \rho^2 \left ( 1 - \sqrt{\frac{r_0}{r}} \right )^{-1}dr^2 \text{ for } r_0\leq r\leq a; \\
\rho^2 = \left ( 1 - \frac{\sqrt{r_0 a}}{r} \right )^{-1} \text{ and } d\rho^2 + dz^2 = \rho^4 dr^2 \text{ for } a\leq r\leq \infty.
\end{cases}
\end{equation}
For the values of $r_0$ and $a$ chosen above, we obtain the surface depicted in Figure~\ref{fig:wormhole}. Notice that now points with $z<0$ correspond to points in a different universe (asymptotically flat region), and not antipodal points in the same universe.

From the plot we may conclude that the exterior region is the same as that in Section~\ref{Schwarzschildsection}, whereas the interior is a flat
cylinder. This arises from the choice of $\Phi = \Phi_0$, which leads to constant $\rho$ for $r_0 \leq r\leq a$. This means that the region inside the wormhole is one of constant redshift.

Regarding motion of light rays, null geodesics correspond to the black straight lines, and thus simply go through the wormhole. On the other hand, geodesics that correspond to the motion of massive particles are those that wind around the surface. These may start in one universe, enter the wormhole and finally leave it through the other side into a different universe. However, if one stops this motion inside the wormhole, then the geodesic becomes a circular one, $dr=0$, and thus the particle will stand still inside the wormhole.

\section*{Conclusions}

In this work we explored static spacetimes by making use of their \textit{Epstein metric}, a Riemannian metric whose geodesics (as we have shown) are in one-to-one correspondence with the causal geodesics of its dual spacetime. The Riemannian nature of the Epstein metric allows a simpler visualization of its geodesics, thus overcoming some of the difficulties inherent to understanding curved spacetimes. Besides this pedagogical aspect, the Epstein correspondence also leads to new interesting results.

We started by studying the Epstein metrics of constant curvature spacetimes, and found that these always led to constant curvature Riemannian manifolds. From these we could easily understand the possible causal geodesics, and consequently the possible types of motion. In the case of the anti-de Sitter spacetime, whose Epstein dual is the sphere $S^4$, we concluded that all geodesics are spatially periodic, that is, it is a Bertrand spacetime.

From the periodicity of the sphere's geodesics we deduced that spherically symmetric spacetimes whose radial sections map to $S^2$ allow radial isochronous oscillatory motion. By starting with a general spherically symmetric spacetime and imposing this condition on its Epstein metric, we determined physically reasonable spacetime metrics with this property when the matter content was a either perfect fluid or an Einstein cluster.

Finally, we studied the radial motion of particles and the motion of light rays in the equatorial plane on some non-constant curvature spacetimes (namely the Schwarzschild, Schwarzschild de Sitter and Schwarzschild anti-de Sitter spacetimes and also a wormhole spacetime) by visualizing the corresponding $2$-dimensional Epstein manifolds. In each case, the most relevant features of the possible motions were apparent in the geometry of the surfaces.

There are still many open questions regarding the Epstein correspondence. For instance, is there an underlying reason why constant curvature spacetimes are mapped to constant curvature Riemannian manifolds? This is not obvious at all, especially because the signs of the curvatures of the two metrics appear to be unrelated. Another prospect for future work is to find other spherically symmetric spacetimes allowing radial isochronous oscillatory motion, possibly using other matter models (like for instance Vlasov). We leave these questions for future research.

\section*{Acknowledgements}
CF gratefully acknowledges the Calouste Gulbenkian Foundation for the scholarship program \textit{Novos Talentos em Matem\'{a}tica}. JN was partially supported by FCT/Portugal through projects UIDB/MAT/04459/2020 and UIDP/MAT/04459/2020 and grant (GPSEinstein) PTDC/MAT-ANA/1275/2014.

\end{document}